\documentclass[reprint,amsmath,amssymb,aps,prl]{revtex4-2}

\usepackage{graphicx}% Include figure files
\usepackage{dcolumn}% Align table columns on decimal point
\usepackage{bm}% bold math
\usepackage{xcolor}
\begin{document}

\title{Thermodynamic performance bounds for radiative heat engines}

\author{Maxime Giteau}
 \email{maxime.giteau.pro@gmail.com}
\author{Michela F. Picardi}%
\author{Georgia T. Papadakis}%
  \email{georgia.papadakis@icfo.eu}
\affiliation{%
 ICFO - The Institute of Photonics Sciences, Castelldefels, Barcelona, Spain
}%

\begin{abstract}
Heat engines cannot generally operate at maximum power and efficiency, imposing a trade-off between the two. Here, we highlight the exact nature of this trade-off for engines that exchange heat radiatively with a hot source. 
We derive simple analytical expressions for the performance bounds of reciprocal and nonreciprocal radiative heat engines. We also highlight that radiative engines can achieve a better power-efficiency trade-off than linear ones.
These bounds are especially relevant for thermophotovoltaics, offering useful metrics against which to compare device performance.
\end{abstract}

%\keywords{Suggested keywords}%Use showkeys class option if keyword
                              %display desired
\maketitle

%\tableofcontents

%\section*{Introduction}

%\textbf{Power-efficiency trade-off}
Heat engines are ubiquitous for energy applications, allowing power generation from any heat source.
Heat engines that can deliver substantial power at high efficiency are earnestly sought after, as any marginal performance increase promises tremendous economic returns due to the large markets involved~\cite{haegel_photovoltaics_2023}.
To optimize the operation of heat engines, it is crucial to understand their performance bounds, particularly the trade-off between power output and efficiency. Thermodynamics, which analyzes energy and entropy flows at the macroscopic level, offers an ideal framework to determine these bounds.

% TPV significance
Radiative heat engines are a class of nonlinear engines that exchange heat as thermal radiation.
One practical implementation of such engines is thermophotovoltaics (TPVs), where a photovoltaic cell directly converts the thermal radiation emitted by a hot source into electricity~\cite{datas_chapter_2021}.
TPVs is a very active and promising research field, with many impressive results reported recently, both in terms of efficiency~\cite{lee_air-bridge_2022,lapotin_thermophotovoltaic_2022} and power output~\cite{lopez_thermophotovoltaic_2023}. 
TPV devices are particularly appealing for energy conversion at very high temperatures (above $1000^\circ \mathrm{C}$, a region where efficiencies well above 50\% are theoretically possible) for applications such as thermal energy storage~\cite{datas_latent_2022}.
Therefore, establishing the general performance bounds of TPV devices is both worthwhile and timely.

% Efficiency bounds
The efficiency of any heat engine operating between a hot source at temperature $T_H$ and a cold sink at temperature $T_C$ is bounded by the Carnot limit $\eta_C = 1 - T_C/T_H$~\cite{greiner_thermodynamics_1995}.
However, except for very particular configurations~\cite{benenti_thermodynamic_2011,buddhiraju_thermodynamic_2018,park_nonreciprocal_2022}, operating close to Carnot efficiency leads to vanishing power output.
In that regard, it is well known that TPV systems can theoretically approach the Carnot efficiency in the limit of zero power output by considering a narrowband emitter~\cite{datas_chapter_2021}.

% Power bounds
At the same time, the power output bounds of radiative heat engines have been explored in great detail, particularly in the context of solar energy conversion, using both detailed balance models~\cite{shockley_detailed_1961,ross_efficiency_1982,giteau_hot-carrier_2022} as well as thermodynamic descriptions~\cite{green_third_2003,vos_reflections_1987,de_vos_endoreversible_1993,landsberg_thermodynamic_1980,ries_complete_1983}.
These bounds readily extend to radiative heat engines by adjusting the temperature of the hot and cold reservoirs.
We emphasize that, as the power received from the sun is free (the sun is outside the conversion system), the term ``efficiency" in the context of solar energy refers to normalized power density. Hence, the quantity being considered and optimized is the power output.
Additionally, the efficiency bound at maximum power, known as the Curzon-Ahlborn limit $\eta_{CA} = 1-\sqrt{T_C/T_H}$ for linear heat engines~\cite{curzon_efficiency_1975,van_den_broeck_thermodynamic_2005}, has been generalized to nonlinear (and particularly radiative) heat engines in the context of endoreversible thermodynamics~\cite{de_vos_efficiency_1985,goktun_design_1993}.
Finally, the efficiency and power output limits of single-, double-, and triple-junction TPV cells have been previously derived using a detailed balance formalism~\cite{datas_optimum_2015}.

% Transition paragraph
Despite these results, the general trade-off between power output and efficiency for radiative heat engines has not been previously investigated.
Furthermore, while universal trade-off relations between power and efficiency have been identified~\cite{shiraishi_universal_2016,pietzonka_universal_2018}, they do not provide specific answers for given classes of heat engines.
As a result, the performance bounds of radiative heat engines (which encompass TPV systems) remain largely unidentified.

%\textbf{This work}
In this paper, we clarify the nature of the power-efficiency trade-off for radiative heat engines, determining the maximum power achievable by a radiative heat engine operating at any given efficiency.
To do so, we generalize the thermodynamic formalism developed for solar energy-conversion limits, first introducing the well-known endoreversible engine model before deriving performance (i.e., power-versus-efficiency) bounds for reciprocal and nonreciprocal engines.

\begin{figure}
\includegraphics[width=\columnwidth]{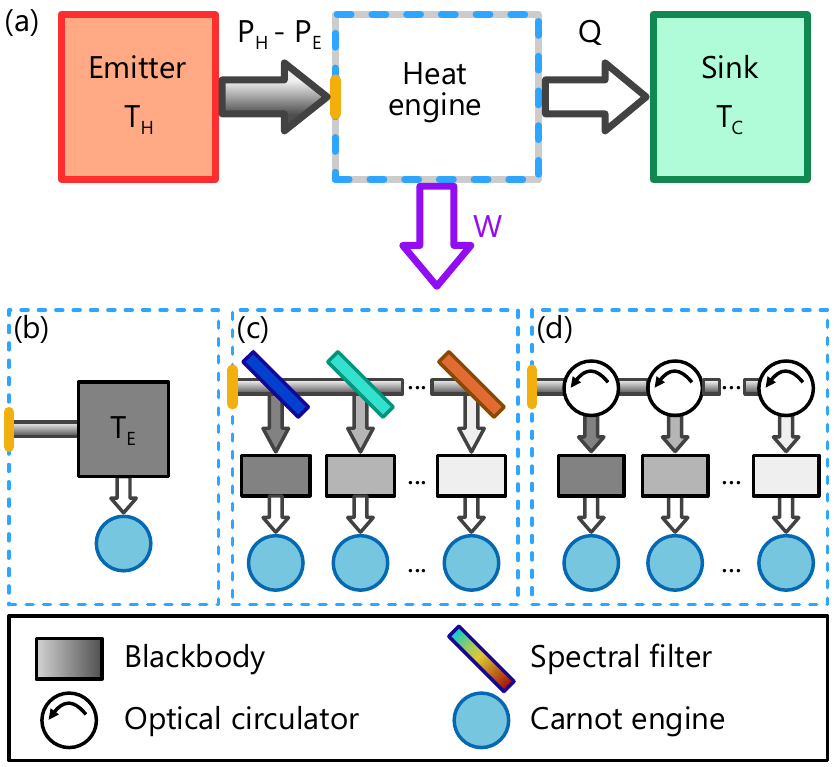}%
\caption{\label{fig:schematic} (a) General representation of a radiative heat engine. (b-d) Different heat-engine models. (b) Endoreversible engine: a single blackbody coupled to a Carnot engine. (c) Infinite reciprocal engine. (d) Infinite nonreciprocal engine for isentropic conversion. 
}
\end{figure}

%\section*{Framework}
We consider a hot emitter at temperature $T_H$ and a cold sink at temperature $T_C$, exchanging power in steady state through a heat engine. The emitter and the engine interact radiatively while the engine is in thermal contact with the sink.
We call $P_H$ and $P_E$ the power densities emitted by the hot emitter and the engine, respectively. 
The radiative heat engine generates an output power density $W$ with an accompanying heat flux $Q$.
The system is schematically represented in Fig.~\ref{fig:schematic}(a).
In the following, we do not account for near-field effects.
We consider that the emitter and heat engine fully see each other (the emitter receives radiation from the heat engine from all directions, and vice versa). As a result, we only need to consider exchanged power \emph{densities} (i.e., power per unit area). In TPVs, this means equating the system efficiency to the pairwise efficiency by assuming a lossless cavity~\cite{burger_present_2020}.

We define a first figure of merit, $\rho$, as the output power density normalized to the power density emitted by a blackbody emitter at temperature $T_H$ (note: this is how \emph{efficiency} is defined for solar energy conversion):

\begin{equation}\label{eq:rho}
    \rho = \frac{W}{\sigma T_H^4},
\end{equation}

\noindent where $\sigma$ is the Stefan-Boltzmann constant. In the following, we refer to this quantity as the power output. Meanwhile, the second figure of merit is the efficiency $\eta$, in which the denominator accounts for the net heat drawn from the emitter (the photons re-emitted towards the emitter do not count as a loss):
\begin{equation}\label{eq:eta}
    \eta = \frac{W}{P_H- P_E}.
\end{equation}

With these definitions, $\rho \leq \eta$, since the net exchanged power must be smaller than the incoming blackbody radiation.
In the following, we consider a blackbody emitter, such that $P_H= \sigma T_H^4$.

%\section*{Blackbody}

First, we present the performance limits of an endoreversible engine consisting of a single blackbody at temperature $T_E$ coupled to a Carnot engine (Fig.~ \ref{fig:schematic}(b))~\cite{de_vos_endoreversible_1993,green_third_2003}. 
Endoreversible means the converter itself is reversible (Carnot engine) while losses arise from the heat exchange with the hot and cold reservoirs.
The power output and efficiency take the expressions:

\begin{equation}\label{eq:rhoBB}
    \rho_E = \left[1-\left(\frac{T_E}{T_H}\right)^4\right]\left[1-\frac{T_C}{T_E}\right]
\end{equation}

\begin{equation}\label{eq:etaBB}
    \eta_E = 1-\frac{T_C}{T_E}.
\end{equation}

From Eqs.~\ref{eq:rhoBB}-\ref{eq:etaBB}, we obtain the \emph{endoreversible model} (power-versus-efficiency) by sweeping the engine temperature $T_E$ from $T_C$ to $T_H$.
Carnot efficiency is achieved in the limit $T_E \to T_H$, leading to zero power output.
The maximum power output $\bar{\rho}_E$ is found by solving $4 T_E^5-3 T_C T_E^4 - T_C T_H^4 = 0$, leading to the so-called blackbody (or endoreversible) limit $\bar{\rho}_E = 85.36 \%$ for solar energy conversion (when $T_H = 6000 \mathrm{K}$ and $T_C=300 \mathrm{K}$)~\cite{vos_reflections_1987,de_vos_endoreversible_1993,green_third_2003}. 
We can also write a direct relation between $\rho_E$ and $\eta_E$, independent of $T_E$:

\begin{equation}
\label{eq:closedform_B}
    \rho_E = \eta_E\left[1-\left(\frac{T_C}{T_H}\right)^4 \frac{1}{(1-\eta_E)^4}\right].
\end{equation}

%Reciprocal and Transition

We emphasize that Eqs.~\ref{eq:rhoBB}-\ref{eq:etaBB} have been derived before, for example in Ref.~\cite{de_vos_endoreversible_1993}. However, they do not impose performance bounds on radiative energy conversion.
To derive general bounds, we consider two cases. In the first, the engine consists of an infinite number of endoreversible subengines, each converting an infinitesimal part of the incident radiation (Fig.~\ref{fig:schematic}(c)). 
This case, detailed in the Supplemental Material~\cite{suppl}, leads to the \emph{reciprocal bound}, \textit{i.e.}, the upper bound for engines that obey time-reversal symmetry. However, as we will see, it only marginally surpasses the endoreversible model.
The second case, presented below, allows for nonreciprocity, leading to the absolute performance bound for radiative energy conversion.
We stress that, for both the reciprocal and nonreciprocal bounds, only power maximization has been previously performed~\cite{landsberg_thermodynamic_1980,green_third_2003} without regard for efficiency.

%\section*{nonreciprocal (absolute) limit}

Here, we consider a radiative heat engine performing isentropic energy conversion, similar to the approach considered by Landsberg for solar energy conversion~\cite{landsberg_thermodynamic_1980,green_third_2003}.
Approaching such an isentropic conversion process in practice requires the radiative heat engine to combine an infinite number of endoreversible subengines connected via nonreciprocal optical components~\cite{ries_complete_1983,green_time-asymmetric_2012,park_reaching_2022} (see Fig.~\ref{fig:schematic}(d)).
Considering an engine temperature $T_E$ and emitted power density $P_E = \sigma T_E^4$, the power output and efficiency are~\cite{green_third_2003}

\begin{equation}\label{eq:rhoN}
\rho_N = 1-\frac{4}{3}\frac{T_C}{T_H}-\left(\frac{T_E}{T_H}\right)^4\left[1-\frac{4}{3}\frac{T_C}{T_E}\right]
\end{equation}

\begin{equation}\label{eq:etaN}
    \eta_N = \frac{\rho_N}{1 - \left(\frac{T_E}{T_H} \right)^4}.
\end{equation}

Eqs.~\ref{eq:rhoN}-\ref{eq:etaN} can be applied to calculate the power output and the efficiency for all engine temperatures $0\leq T_E \leq T_H$, leading to the \emph{nonreciprocal bound} for radiative energy conversion.
The power output is maximized for $T_E=T_C$, leading to:
\begin{equation}\label{eq:rhobarN}
    \bar{\rho}_N = 1- \frac{4}{3}\frac{T_C}{T_H}+\frac{1}{3}\left(\frac{T_C}{T_H}\right)^4,
\end{equation}

\noindent which gives $\bar{\rho}_N = 93.33 \%$ for $T_C = 300 \ \mathrm{K}$ and $T_H = 6000 \ \mathrm{K}$, the absolute (Landsberg) limit for solar energy conversion~\cite{landsberg_thermodynamic_1980,green_third_2003}.
In the limit $T_E \to T_H$, the efficiency tends to the Carnot limit $\eta_C$ as the power output goes to 0.
We note that operating at $T_E \leq T_C$ is always suboptimal as both efficiency and power can increase from that point.
By combining Eqs.~\ref{eq:rhoN} and \ref{eq:etaN}, we can also write a closed-form expression between $\eta_N$ and $\rho_N$, independent of $T_E$:

\begin{equation}
\label{eq:closedform_N}
    \eta_N = 1 - \frac{4}{3}\frac{\eta_N}{\rho_N}\frac{T_C}{T_H}\left[1-\left(1-\frac{\rho_N}{\eta_N}  \right)^{3/4} \right].
\end{equation}

We have thus obtained simple analytical expressions for the universal thermodynamic bound of radiative energy conversion between bodies at temperatures $T_H$ and $T_C$. A blackbody emitter is optimal as it offers the highest incident power and, therefore, the highest power output for any given efficiency.

%\section*{Illustrations}

\begin{figure}
\includegraphics[width=\columnwidth]{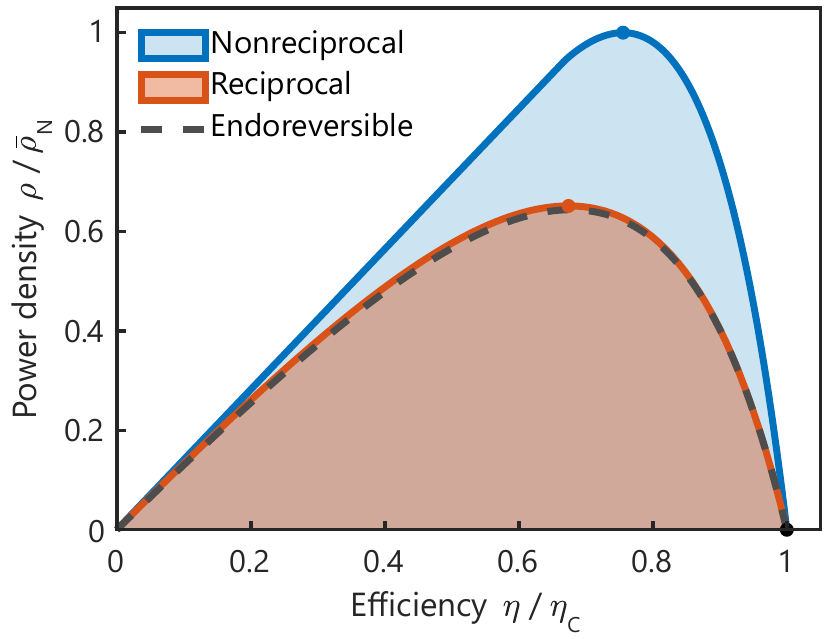}%
\caption{\label{fig:fig2} Thermodynamic performance bounds for radiative energy conversion: maximum power output $\rho$ (normalized to the nonreciprocal limit $\bar{\rho}_N$) as a function of the efficiency $\eta$ (normalized to the Carnot efficiency $\eta_C$) for the nonreciprocal bound (blue area), the reciprocal bound (red area) and the endoreversible engine model (dashed black line).
The emitter temperature is $T_H = 600 \ \mathrm{K}$ and the sink temperature is $T_C = 300 \ \mathrm{K}$.}
\end{figure}

We compare in Fig.~\ref{fig:fig2} the power-versus-efficiency bounds for the endoreversible model (Eqs.~\ref{eq:rhoBB}-\ref{eq:etaBB}), the reciprocal bound (see Supplemental Material~\cite{suppl}) and the nonreciprocal bound (Eqs.~\ref{eq:rhoN}-\ref{eq:etaN}), considering an emitter temperature $T_H = 600 \ \mathrm{K}$ and a sink temperature $T_C = 300 \ \mathrm{K}$.
We note that these bounds are only a function of the temperature \emph{ratio} $T_C/T_H$, as seen from the closed-form expressions (Eqs.~\ref{eq:closedform_B} and \ref{eq:closedform_N}).
Crucially, the reciprocal bound only marginally surpasses the endoreversible model. Since it can be calculated very simply (Eq.~\ref{eq:closedform_B}), the endoreversible engine model is, therefore, a useful approximation for the upper bound of reciprocal systems, and we will use both interchangeably in the following.

\begin{figure}
\includegraphics[width=\columnwidth]{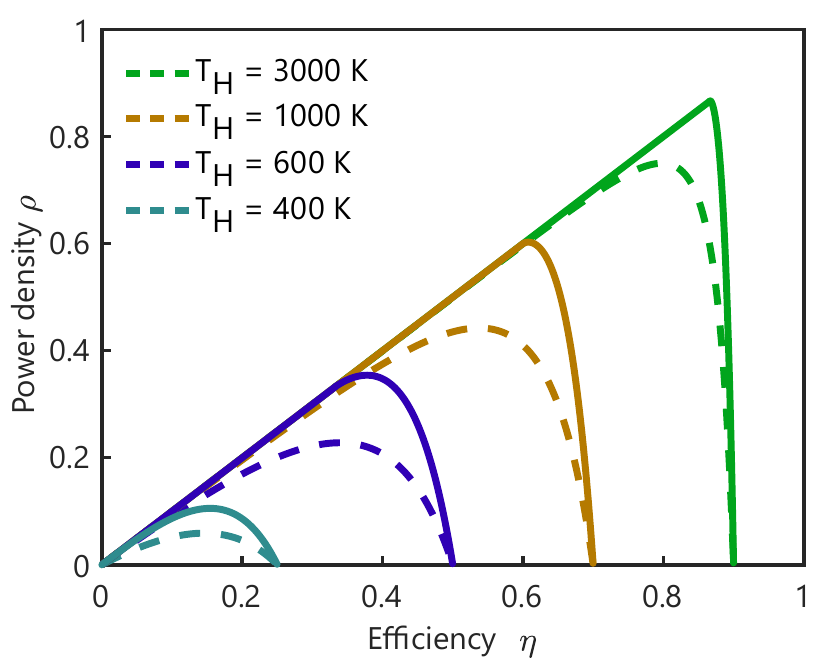}%
\caption{\label{fig:fig3} Nonreciprocal (solid lines) and reciprocal (dashed lines) performance bounds in terms of maximum power output $\rho$ as a function of the efficiency $\eta$ for different emitter temperatures. The sink temperature is $T_C = 300 \ \mathrm{K}$.}
\end{figure}

Next, we show in Fig.~\ref{fig:fig3} the nonreciprocal and reciprocal bounds for different emitter temperatures.
We observe that the relative difference between both bounds tends to be more significant as the emitter temperature decreases or when the efficiency approaches the Carnot limit (see Supplemental Material~\cite{suppl} for a figure with their ratio).
Furthermore, we can show that the endoreversible model and the nonreciprocal bound are bounded by

\begin{align}
    \rho_E &\leq 4 \frac{T_H}{T_C} \eta_E (\eta_C-\eta_E) \label{eq: rhoB_quad} \\
    \rho_N &\leq 8 \frac{T_H}{T_C} \eta_N (\eta_C-\eta_N),\label{eq: rhoN_quad}
\end{align}
\noindent and that these bounds are approached as $T_E \to T_H$ (first-order approximation). 
Therefore, introducing nonreciprocity enables twice as much power output when operating close to the Carnot limit.
We also emphasize that Eqs.~\ref{eq: rhoB_quad}-\ref{eq: rhoN_quad} follow the quadratic form of the ``universal" bounds derived in previous works~\cite{shiraishi_universal_2016,pietzonka_universal_2018}.

\begin{figure} [ht]
\includegraphics[width=\columnwidth]{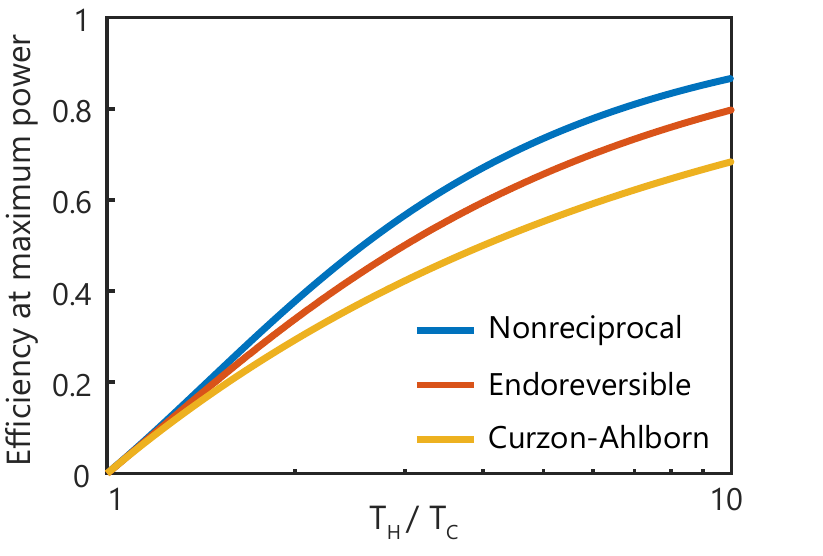}%
\caption{\label{fig:fig4} Efficiency at maximum power for different bounds as a function of the temperature ratio $T_H/T_C$.}
\end{figure}

Last, we focus on the trade-off between power output and efficiency in radiative heat engines by comparing their efficiency at maximum power to that of linear heat engines.
For linear heat engines, the Curzon-Ahlborn limit, originally derived for endoreversible heat engines~\cite{curzon_efficiency_1975}, was shown to extend to all irreversible linear engines~\cite{van_den_broeck_thermodynamic_2005}. For nonlinear heat engines, efficiency at maximum power has been previously derived, albeit only for endoreversible engines~\cite{de_vos_efficiency_1985}.
When $T_H$ tends to $T_C$, the behavior of radiative heat engines becomes linear, so their efficiency at maximum power approaches the Curzon-Ahlborn limit $\eta_{CA} \approx \eta_C/2$.
Beyond the linear approximation, we show the efficiency at maximum power for the endoreversible model and the reciprocal bound in Fig.~\ref{fig:fig4}.
We observe that the efficiency at maximum power for a given temperature ratio $T_H/T_C$ follows $\bar{\eta_N} \geq \bar{\eta_E} \geq \eta_{CA}$.
This result, which originates from nonlinearity~\cite{de_vos_efficiency_1985}, highlights that radiative heat engines such as TPV cells can achieve a better trade-off between power output and efficiency than conventional linear heat engines, particularly for large temperature ratios.
Furthermore, for nonlinear engines such as radiative heat engines, nonreciprocity can be exploited to achieve performance beyond endoreversible thermodynamics.

%\section*{Conclusion}

Despite their broad validity, the bounds derived here may be overcome in two cases. The first is near-field operation, where the heat transferred from the emitter to the engine can be much larger than in the far field~\cite{laroche_near-field_2006}. The second is the introduction of a heat engine on the hot side, either to increase the power radiated to the cold side (thermophotonics~\cite{harder_thermophotonics_2003}) or to generate additional power output on the hot side (thermoradiative cells~\cite{strandberg_theoretical_2015,buddhiraju_thermodynamic_2018,park_nonreciprocal_2022}).

In conclusion, we derive thermodynamic performance bounds for radiative energy conversion, expressed in terms of the maximum power achievable for any efficiency.
In doing so, we integrate the limits derived over the previous decades for solar energy conversion within a larger thermodynamic framework.
The nonreciprocal bound, which has a simple analytical expression, establishes the universal performance limit of radiative heat engines.
At the same time, the endoreversible engine model offers a good approximation for the bound of reciprocal engines.
We show that nonreciprocal systems may significantly outperform reciprocal ones, especially for low emitter temperatures or when operating at high efficiency.
We also reveal that radiative heat engines may achieve a better power-efficiency trade-off than conventional linear engines, especially for high-temperature sources.
These results add to the attractive potential of TPV systems for high-temperature energy conversion and offer useful figures of merits and bounds for their comparison.

\section*{Acknowledgments}

M.G. would like to thank Daniel Suchet for stimulating discussions.
This work has been supported in part by the ``la Caixa" Foundation (ID 100010434), the Spanish MICINN (PID2021-125441OA-I00, PID2020-112625GB-I00, and CEX2019-000910-S), the European Union (fellowship LCF/BQ/PI21/11830019 under the Marie Skłodowska-Curie Grant Agreement No. 847648), Generalitat de Catalunya (2021 SGR 01443), Fundació Cellex, and Fundació Mir-Puig.
M.G. and M.F.P. acknowledge financial support from the Severo Ochoa Excellence Fellowship.

\bibliography{biblio.bib}

\end{document}

\title{Thermodynamic performance bounds for radiative heat engines \\ Supplemental Material}

\author{Maxime Giteau}
\author{Michela F. Picardi}%
\author{Georgia T. Papadakis}%
\affiliation{%
 ICFO - The Institute of Photonics Sciences, Castelldefels, Barcelona, Spain
}%

\maketitle

\section*{Reciprocal bound}

We derive here the performance bounds of a reciprocal radiative heat engine by analogy with the multicolor limit for solar energy conversion~\cite{green_third_2003}. We assume the heat engine consists of an infinite set of endoreversible sub-engines at temperatures $T_E(E)$, each converting the radiation from a spectral element $[E, E+dE]$ at Carnot efficiency, as in Fig.~1(c). We define the photon occupation numbers

\begin{align}
    n_H (E) &= \frac{1}{\exp[E / k T_H]-1} \\
    n_E (E) &= \frac{1}{\exp[E / k T_E(E)]-1},
\end{align}
\noindent as well as $\displaystyle \Delta n (E) = n_H(E) - n_E(E)$.
The power generated by each endoreversible engine is

\begin{equation}\label{eq:dPR}
    dW = \frac{2 \pi}{h^3 c^2} E^3  \Delta n (E) \left[1- \frac{T_C}{T_E(E)} \right] dE.
\end{equation}

By summing the power generated by all engines, i.e., by integrating $dW$, we obtain

\begin{equation}\label{eq:rhoR}
    \rho_R = 1 - \frac{15}{\pi^4 k^4 T_H^4} \int_0^\infty E^3 \left[ \frac{T_C}{T_E(E)} \Delta n (E)
    + n_E(E) \right] dE
\end{equation}

\begin{equation}\label{eq:etaR}
    \eta_R = 1 - \frac{ \int_0^\infty E^3 \frac{T_C}{T_E(E)} \Delta n (E) dE }
    { \int_0^\infty E^3 \Delta n (E) dE}.
\end{equation}

\noindent where $k$ is the Boltzmann constant.
For a given $\eta_R$, $T_E$ functions must satisfy, according to Eq.~\ref{eq:etaR}:

\begin{equation}
\label{eq:T_FtoSatisfy}
    \int_0^\infty E^3 \left[\frac{1}{T_E(E)} - \frac{1}{T_F} \right] \Delta n (E) dE = 0,
\end{equation}
%\begin{equation}\label{eq:etaEqualTF}
%    \int_0^\infty E^3 \left[\frac{1}{T_E(E)} - \frac{1}{T_F} \right] \Delta n (E) dE = 0,
%\end{equation}
\noindent where $\displaystyle T_{F} = \frac{T_C}{1-\eta_R}$, verifying $T_C \leq T_F \leq T_H$.
A trivial solution is for all blackbodies to be at the same temperature $T_E = T_F$, leading to the endoreversible model presented in the main text.
However, this is not the upper bound for reciprocal radiative energy conversion, as the temperature profile can be refined to improve performance.

We now focus on determining the temperature profile $T_E(E)$ that maximizes the power output for any given efficiency.
First, we calculate the absolute maximum power for reciprocal radiative energy conversion by finding the temperature $T_E$ that maximizes $dW$ (Eq.~\ref{eq:dPR}) for each photon energy.
%, by solving:
%\begin{equation}
   %\Delta n = \frac{E}{k T_C}\left( 1-\frac{T_C}{T_E} \right) \frac{\exp[E/k T_E]}{(\exp[E/k T_E]-1)^2}
%\end{equation}
%alternative form with fewer $T_E$'s:
%\begin{equation}\label{eq:TCMC}
%    \frac{\exp[E / k T_E]-1}{\exp[E/k T_H]-1} = 1 + \frac{E}{k T_C}\left( 1-\frac{T_C}{T_E} \right)\left(1+ \frac{1}{\exp[E/k T_E]-1}\right)
%\end{equation}
%In the limit $E \to 0$, the temperature tends to $T_E = \sqrt{T_C T_H}$.
For solar energy conversion, we recover the so-called multicolor limit $\bar{\rho}_R = 86.82\%$~\cite{green_third_2003}.
More generally, from Eqs.~\ref{eq:rhoR}-\ref{eq:etaR}, we find that, for a given efficiency $\eta_R$, maximizing the power density is equivalent to minimizing the emitted power density

\begin{equation}
    P_E \propto \int_0^\infty E^3 n_E(E) dE.
\end{equation}

%We note that a blackbody emitter is optimal since suppressing energy transfer at any energy $E$ can equivalently be achieved by setting $T_E(E) = T_H$.

\begin{figure} [h]
\includegraphics[width=\columnwidth]{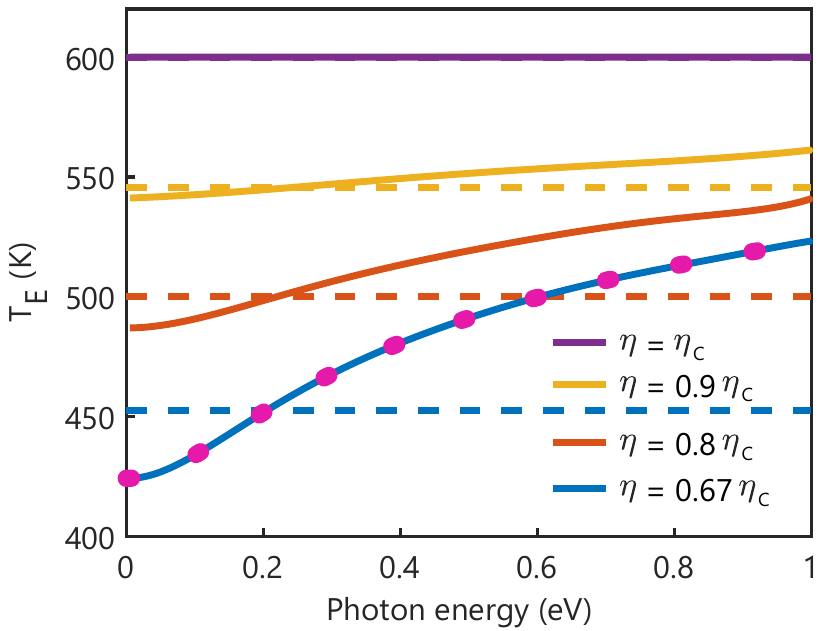}%
\caption{\label{fig:figS1} Engine temperature $T_E$ as a function of the photon energy for the reciprocal bound (solid lines) compared to the fixed temperature of the endoreversible model (dashed lines) for different efficiency points. The blue line corresponds to the temperature profile for the maximum power point $\bar{\rho}_R$. We consider an emitter temperature $T_H= 600 \ \mathrm{K}$ and a sink temperature $T_C = 300 \mathrm{K}$. The pink dots show the point-by-point optimization that leads to maximum power.}
\end{figure}

To find the optimal solutions, we model $T_E(E)$ as a polynomial and adjust its coefficients to minimize $P_E$ while satisfying Eq.~\ref{eq:T_FtoSatisfy}.
We show in Fig.~\ref{fig:figS1} (solid lines) the temperature profiles $T_E(E)$ that lead to the reciprocal bound for several efficiencies for the conditions considered in Fig.~2 of the article ($T_H= 600 \ \mathrm{K}$ and $T_C= 300 \ \mathrm{K}$). 
These temperature profiles are obtained by fitting the coefficients of a 7th-order polynomial to maximize the power for a given efficiency.
We compare these profiles to the fixed engine temperatures $T_E=T_F$ for the endoreversible model (dashed lines). 

We observe that the function $T_E$ crosses $T_F$ for a photon energy slightly above 0.2 eV (close to the blackbody emission peak at 600 K). 
The maximum power obtained by polynomial fitting (blue line) perfectly matches the energy-by-energy maximization of the output power (pink dots), validating the optimization approach.

\section*{Power ratio between nonreciprocal and reciprocal bounds}

\begin{figure} [h]
\includegraphics[width=\columnwidth]{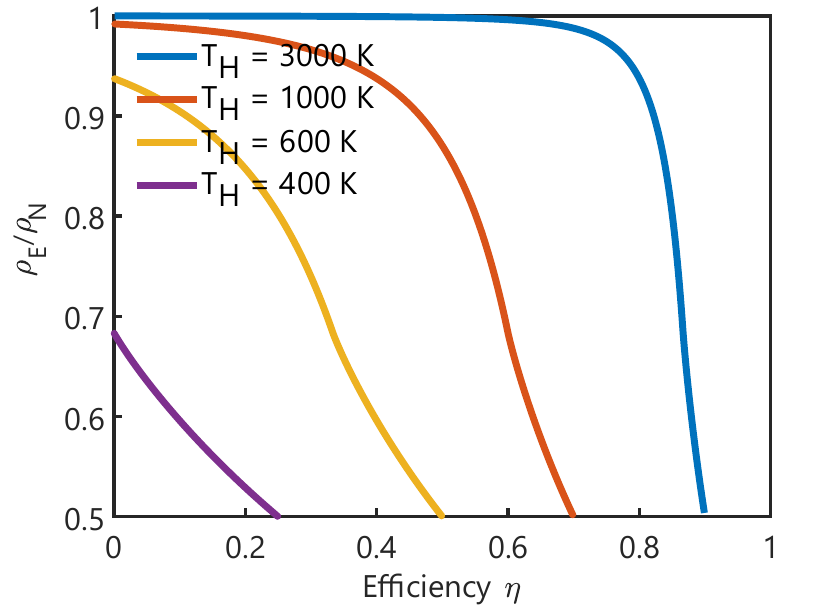}%
\caption{\label{fig:figS2} Ratio between the power outputs for the endoreversible model $\rho_E$ and the nonreciprocal bound $\rho_N$ as a function of the efficiency $\eta$, for different emitter temperatures. The sink temperature is $T_C = 300 \ \mathrm{K}$. Close to the Carnot efficiency limit, nonreciprocal engines generate twice as much power as reciprocal ones.}
\end{figure}

\bibliography{biblio.bib}